\documentclass[12pt]{revtex4}
\usepackage{amsmath,amssymb,times}
\usepackage{dcolumn}
\usepackage{bm}
\usepackage{color}
\usepackage{amsthm,amscd,mathrsfs}
\usepackage{graphicx}
\theoremstyle{definition}

\def\M{{{M}}}
\def\F{{{{F}}}}
 
\def\pp{{\parallel}}
\def\D{{\Delta}}

\begin{document}
\title{Finsler Geometrical Path Integral}
\author{Takayoshi Ootsuka}
 \email{ootsuka@cosmos.phys.ocha.ac.jp}
\author{Erico Tanaka}
  \email{erico.tanaka01@upol.cz}
  \affiliation{$^*$Physics Department, Ochanomizu University, 2-1-1 Ootsuka Bunkyo Tokyo, Japan }
  \affiliation{$^\dagger$Mathematics Department, Palacky University, Svobody 26, Olomouc, Czech Republic}
  \affiliation{Advanced Research Institute for Science and Engineering,
Waseda University, 3-4-1 Ohkubo Shinjuku, Tokyo, Japan}
\begin{abstract}
A new definition for the path integral is proposed in terms of Finsler geometry. 
The conventional Feynman's scheme for quantisation by Lagrangian formalism 
suffers problems due to the lack of geometrical structure of the 
configuration space where the path integral is defined. 
We propose that, by implementing the Feynman's path integral on 
an extended configuration space endowed with a Finsler structure, 
the formalism could be justified as a proper scheme for quantisation from Lagrangian only,
that is, independent from Hamiltonian formalism. 
The scheme is coordinate free, and also a covariant framework which does not depend on 
the choice of time coordinates.
\end{abstract}
\maketitle
\section{Introduction}
Feynman himself, stated that fundamentally there are no new results, 
when he first proposed the quantisation by Lagrangian formalism,  
in other words, the path integral formulation~\cite{Feynman1}. 
Now, after more than 60 years, its impact and usefulness cannot be overestimated, 
especially after the invention of Feynman diagram and its application to 
covariant perturbation theory. 
However, the appealing point of this formulation is not just practical calculations, 
but the basic ideas stuffed in its foundation, 
which give us a different perspective 
from the comparatively well established canonical quantisation; 
and also suggest us the possibility to be the coordinate free and covariant form 
of quantum theory. 
The central core philosophy of path integral is the belief in variational principle. 
By this principle, the classical path is chosen out from 
infinitely many paths by ``variation'', and it is only this classical path which is realised. 
Therefore, 
we expect that there exists a more fundamental theory, 
which serves as a basis for the classical theory 
i.e., the quantum theory. 
Constructing the path integral means that 
we take the inverse operation of variation, 
and try to reach the fundamental quantum theory from classical mechanics. 
This is easy in words but difficult to realise, 
in fact, the original formulation by Feynman lacks rigorous mathematical description, 
which prevents the formulation to be 
``a third formulation of quantum theory'', as Feynman stated. 
That is to say, the formulation is not self-contained, 
and the defect could be easily recognised by calculating concrete examples. 
For instance, when one tries to use general curvilinear coordinates 
or particle constrained on a certain surface, 
naive application of Feynman's original formula does not work. 
The well-known resolution to obtain correct calculational results in above cases 
is to use the Hamiltonian formalism auxiliary. 
The reason why this technique is efficient is that, 
phase space which is the stage for Hamiltonian formalism 
is a symplectic manifold, and there exists a geometrical object; a symplectic form.
Symplectic form defines a canonical volume, a Liouville measure 
on the phase space. 
In contrast, the original Feynman's path integral for 
a quantum mechanical particle is defined on a 
$\mathbb{R}^3$ Euclidean space. 
In this Euclidean space, there is a Euclidean measure naturally defined from the Euclidean 
structure. However, even this measure could not be used without an adjustment of the 
factor $1/A$, as described in Feynman's paper~\cite{Feynman1}.  
The intervention to Hamiltonian formalism 
has the effect of covering this defect, 
it compensates the lack of geometrical structure of configuration space, 
by the geometrical structure of symplectic manifold. 
In summary, path integral had never been an independent nor robust 
quantisation by Lagrangian formalism, 
and the main cause is its lack in geometrical setting. 
Due to this defect, in principle, 
being provided with an arbitrary Lagrangian 
is insufficient for this formulation to work. 

In this letter, we will try to construct a true quantisation scheme by Lagrangian 
formalism, by faithfully following the philosophy of path integral Feynman proposed. 
Since the conventional configuration space has no 
geometric structure that could be used as a stage of geometry, 
we consider the extended configuration space 
that could be canonically endowed with a Finsler structure 
determined from the Lagrangian. 
We will take this Finsler geometry 
for the backbone of our formulation. 
To distinguish from the conventional path integral, let us call this a 
{\it Finsler geometrical path integral}, or for short, just {\it Finsler path integral}. 
The formulation is geometrical by construction, therefore, 
its covariance and coordinate independence could be easily verified, 
and the problems that conventional method suffers will be solved automatically. 

\section{Finsler Geometry}
Finsler geometry, which is a generalisation of Riemannian geometry, 
has been given relatively small attention by physicists in spite of its 
wide potential ability of describing physical applications. 
This seems mainly because of its calculational complexity.
Our approach taken in this letter does not require any expression of 
line elements nor non-linear connections, 
which are the major source of complexity. 
Following Tamassy~\cite{Tamassy1}, we emphasise that the Finsler manifold 
we are referring to as a ``point Finsler space'', 
and refrain from the concept of ``line element Finsler space'', 
proposed by Cartan, though the latter is usually regarded as the 
standard approach. 
For our motivation, the former is a more simple approach, 
and we also expect it to be more appropriate for 
further physical applications.

Finsler manifold $({\M} , {\F})$ is a set of differentiable manifold $\M$ 
and a Finsler structure 
$\F:v \in T_x\M \rightarrow F(x,v) \in \mathbb{R}$, 
obeying the following homogeneity condition:
\begin{eqnarray}
\F \left({ x},\lambda {v}\right) 
= \lambda \F\left({x},{v}\right) \quad \lambda > 0, \, \, x \in {M},
 \,\, v \in T_x{M}.  \label{finslerfunc}
\end{eqnarray}
$\F$ gives the distance for the oriented curve on $\M$.  
Taking a parametrisation $t$, 
The length of a curve $C$ is given by 

\begin{eqnarray}
\eta[C] = \int_C {\F} = \int_a^b F\left({x}(t),\frac{d{x}(t)}{dt}\right)dt.
\end{eqnarray}

$d\eta = {\F}({x},{dx})$ is the infinitesimal distance between 
two points ${ x}$ and ${ x}+dx$.
$\eta[C]$ depends on the orientation of the curve $C$, 
but by the homogeneity condition, it does not depend on the 
parametrisation of $C$. 
The Riemannian geometry is a special case for Finsler geometry when 
$\F({x},d{ x}) = \sqrt{g(dx, dx)}$, where $g$ is a Riemannian metric. 
Relativistic particle is also an important example of Finsler geometry. 
Moreover, we emphasise that Lagrangian mechanics could also be regarded as 
Finsler geometry by 
considering the extended configuration space $M = \mathbb{R} \times Q^n$ 
instead of the configuration space $Q^n$, 
together with a Finsler function $\F$ which is given by the relation, 
\begin{eqnarray}
&&{\F}\left(x^0, x^1, \cdots, x^n, dx^0, dx^1, \cdots ,dx^n\right)  \nonumber \\ 
&&= L \left(x^1, \cdots, x^n,  \frac{dx^1}{dx^0}, \cdots ,\frac{dx^n}{dx^0},x^0\right)|dx^0|, 
\end{eqnarray} 
when the Lagrangian $L(x,\dot{x},t)$ is provided. 
Then $(\M,\F)$ forms a Finsler manifold. 

Throughout this letter we require only the above homogeneity condition for $\F$, 
and we will {\it not} use the 
metric $g_{ij}=\frac12\frac{\partial^2 F^2}{\partial y^i\partial y^j}$, 
which is the standard object in the ``line element Finsler space''
~\cite{Cartan1, Chern2}. 

\section{Indicatrix, indicatrix body and area}
Let $(\M,\F)$ be a Finsler manifold, 
and ${\rm dim} \M = n+1$.
Suppose we have a $k$-dimensional submanifold $\Sigma$ of $M$, 
and $x$ be the point on $\Sigma$. 
Tangent space of $\Sigma$ at point $x$ is denoted by $T_x \Sigma$. 
The definition of ``area'' in Finsler manifold, i.e., 
a measure of ${\Sigma}$ with $k \leq n+1$ is given 
by Busemann and Tamassy~\cite{Busemann2, Tamassy1}, 
using indicatrix and indicatrix body. 
Indicatrix is a ruler which measures a ``unit" area by means of 
a Finsler function, and indicatrix body is the domain cut out by the indicatrix. 
The definition of indicatrix is given by, 

\begin{eqnarray}
I_x := \{ {v} \in T_x {M} | F ({x},{v}) = 1 \}
\end{eqnarray}
and indicatrix body by, 
\begin{eqnarray}
D_x := \{ {v} \in T_x {M} | F ({x},{v}) \leq 1 \}.
\end{eqnarray}
 
For the Riemannian geometry, a special case of Finsler geometry, 
the indicatrix becomes a quadric surface. 
Busemann and Tammasy proposed that, 
the Finsler area (measure) could be defined by setting the 
ratio of two domains, $\D \Sigma_x$; 
the infinitesimally small element of surface tangent to $\Sigma$ at point $x$, 
and $ T_x \Sigma \cap D_x$; the intersection of the tangent space of $\Sigma$
with indicatrix body, to
\begin{eqnarray}
\pp \D \Sigma_x \pp_F : \pp T_x \Sigma \cap D_x \pp_F 
= \pp \D \Sigma_x \pp_R : \pp T_x \Sigma \cap D_x \pp_R, \label{eq.area}
\end{eqnarray}
using an appropriate Riemannian structure.
$\pp \,\pp_F$ and $\pp \,\pp_R$ denotes the measure defined on 
Finsler manifold and Riemannian manifold, respectively. 
Note that since it is a ratio, this value does not depend on the 
choice of Riemannian structure. 

\begin{figure*}[t]
  \begin{center}
   \includegraphics[width=90mm]{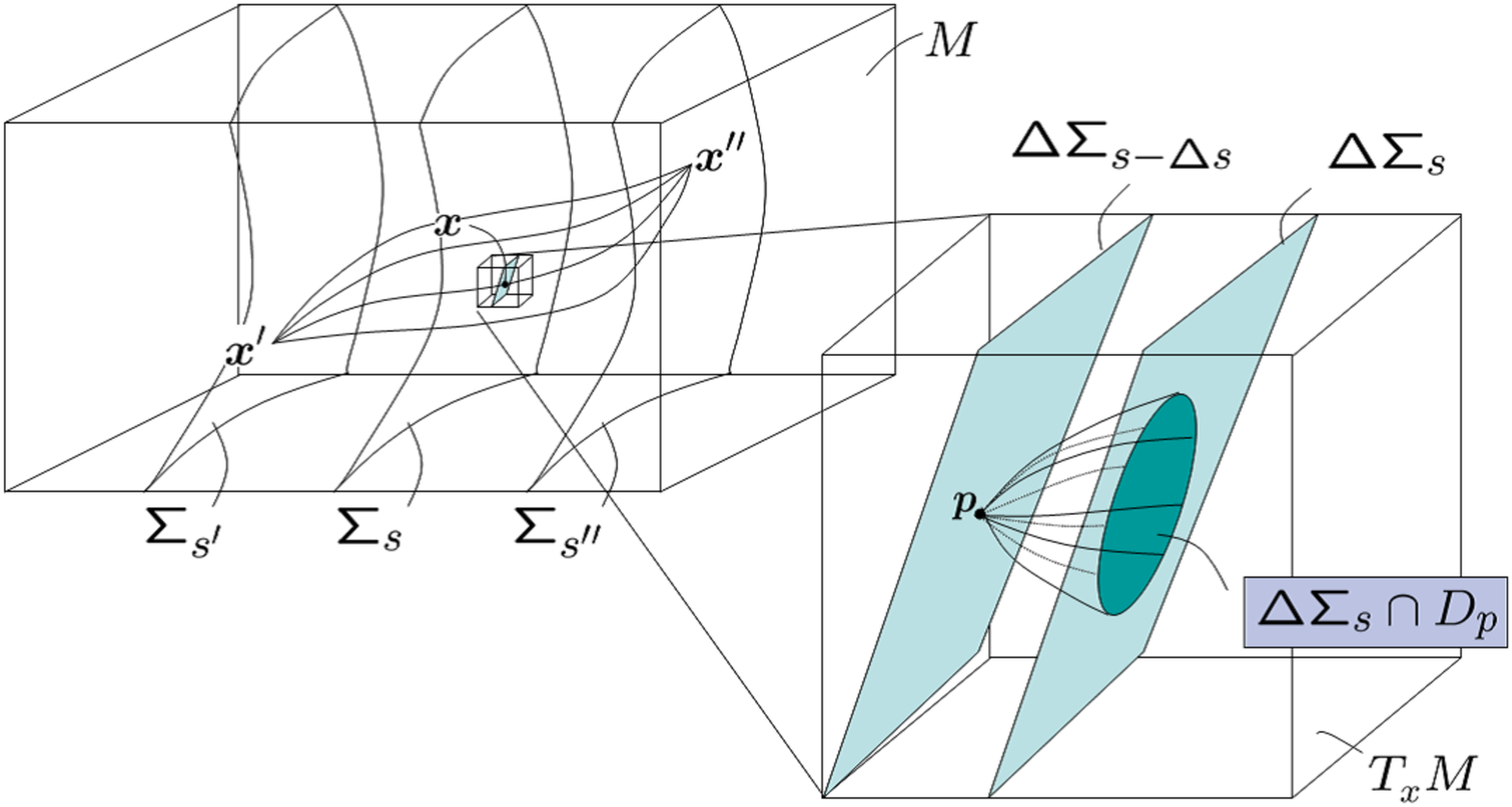}
   \end{center}
  \caption{  }
  \label{fig1}
\end{figure*}

However, while Busemann and Tamassy considered Finsler manifold where its 
indicatrix body was compact, 
for our case of physics, 
in general it would be non-compact 
and also the neighbourhood of point ${ y} = 0$ is not contained. 
So, there are cases such that, $T_x \Sigma \cap D_x = \phi$. 
We need to define a measure 
also applicable for these cases. 
We propose the following definition by using the perspective of path integral itself. 
We will only consider the area for $k=n$ case, i.e. a Finsler area of a hypersurface. 
Assume that there exists a foliation satisfying the following condition: 
i) choose initial point $x'$ and final point $x''$ from two 
different leaves, such that these points can be connected by curves and 
on this curve ${F}(x,dx)$ is well-defined. ii) The 
leaves of foliation are transversal to these set of curves.  
FIG.\ref{fig1} shows such foliation in a simplified way. 
($M$ is figured as a rectangular parallelepiped just for visibility.) 
The leaves; hypersurfaces $\Sigma$, 
are labelled by parameter $s$ which is a function on $\M$, and represents the time variable. 
Consider a tangent space $T_x M$, where $x \in \Sigma_{s}$, 
and denote by $\Delta \Sigma_s$, 
the infinitesimally small element of surface tangent to $\Sigma_{s}$ at point $x$. 
Taking infinitesimally small $\D s$, 
a slightly dislocated hyperplane $\Delta \Sigma_{s-\Delta s}$ would be 
also included in $T_x M$, parallel to $\D \Sigma_s$. 
Take an arbitrary point $p$ from $\Delta \Sigma_{s-\Delta s} \cap T_x \M$. 
Then the indicatrix body at point $p$ 
could have an intersection with $\Delta \Sigma_s$, 
as shown in FIG.\ref{fig1}. 
Then, $\D\Sigma_{s- \D s} \cap D_p \neq \phi$, and 
we can generalise the formula (\ref{eq.area}) to 
\begin{eqnarray}
\pp \Delta \Sigma_s \pp_F 
=\omega \lim_{\Delta s \to 0}
\frac{\pp \Delta \Sigma_s \pp_R}{ \pp \Delta \Sigma_s\cap D_{p} \pp_R }, \label{eq.area2}
\end{eqnarray}
which gives the same result as Tamassy's measure for Riemannian case. 
In general, 
$\omega = \pp \Delta \Sigma_s\cap D_{p} \pp_F$ 
could not be decided by geometry only. 
However, it could be determined by the condition that the propagator approaches a delta function 
in the limit of $\Delta s \rightarrow 0$, which we now assume to be a constant. 
The important thing is that the measure above is defined 
without any introduction of Euclidean space or 
Euclidean structure, 
and still it is capable of giving the correct measure up to an irrelevant constant factor. 

\section{Finsler path integral}
Now the Finsler path integral could be defined as,  
\begin{eqnarray}
\mathscr{U}[\Sigma_{s''}, \Sigma_{s'}] 
&=&\int_{s'}^{s''} 
\hspace{-6pt}
\delta C~{\rm e}^{\frac{i}{\hbar}\int_{C} F} \nonumber \\
&=& \lim_{\Delta s \to 0}d \Sigma_{s'}
\int_{\Sigma_{s_1}} 
d\Sigma_{s_1} \cdots 
\int_{\Sigma_{s_N}} 
d\Sigma_{s_{{N-1}}} 
\exp \left(\frac{i}{\hbar} \sum_{j={0}}^{{N-1}} 
\eta[\gamma_{x_j}^{x_{j+1}}]
\right). \label{eq.finslerPI}
\end{eqnarray}
Here, $d\Sigma_{s'}, d\Sigma_{s_1}, \cdots d\Sigma_{s_{{N-1}}}$ are 
the previously defined Finsler measure, 
$\gamma_{x_j}^{x_{j+1}}$ a geodesic 
connecting the point $x_j$ on $\Sigma_{s_j}$  and $x_{j+1}$ on $\Sigma_{s_{j+1}}$, 
and $\eta[\gamma_{x_j}^{x_{j+1}}]$ 
is the Finsler length of $\gamma_{x_j}^{x_{j+1}}$. 
Since our definition of path integral stands on pure 
geometrical construction and the geodesic 
$\gamma$ only depends on Finsler structure, 
it is a coordinate free formulation. 
The evolution of the Schr\"odinger function is given by this propagator by, 
\begin{eqnarray}
\psi_{\Sigma_{s''}} 
= \int_{\Sigma_{s'}} \mathscr{U}[\Sigma_{s''}, \Sigma_{s'}]\psi_{\Sigma_{s'}}, \label{eq.prop}
\end{eqnarray}
where $\psi_{\Sigma_s}$ is a Schr\"odinger's wave function 
defined on $\Sigma_s$.
It also points out that the wave function could be 
only realised on the leaves of foliation. 
Note that unlike the usual propagator, 
there is a $n$-form $d\Sigma_{s'}$ in (\ref{eq.finslerPI}), 
which makes 
$\psi_{\Sigma_{s''}}$ a function ($0$-form) on $\Sigma_{s''}$ by integration. 

\section{Examples} \label{sec.examples}
\begin{figure*}[htbp]
  \begin{center}
   \includegraphics[width=50mm]{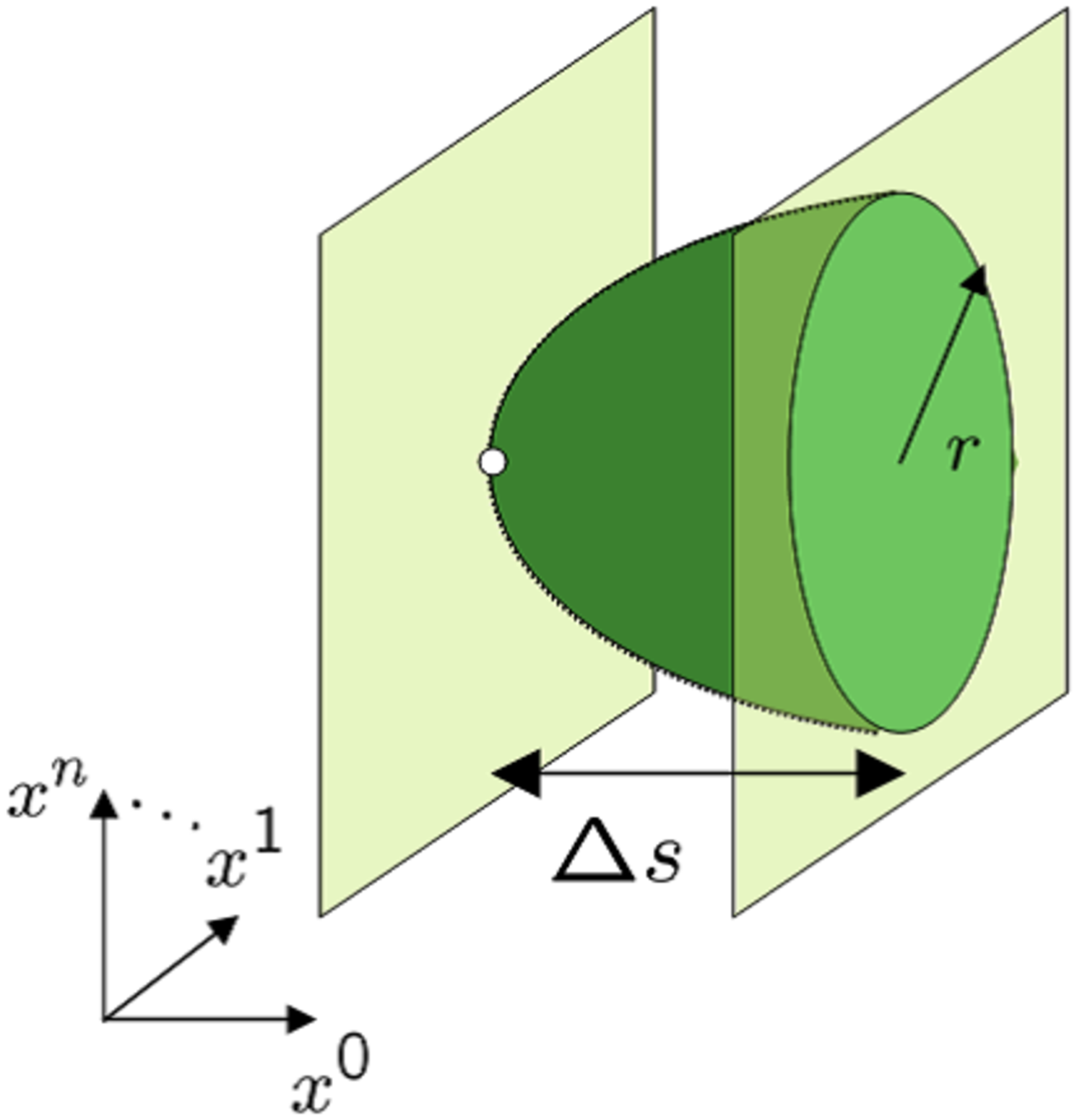}
  \end{center}
  \caption{ }
  \label{fig2}
\end{figure*}

Here we introduce several applications that prove the validity and effectiveness 
of the proposed formula. 
We first calculate the simplest example 
for a non-relativistic particle moving in a potential $V$. 
Consider the Finsler manifold $(M,F)$ with ${\rm dim} M = n+1$, 
and Finsler function defined by 
\begin{eqnarray}
{F}\left(x^0, x^i, dx^0, dx^i\right) = \frac{m}{2}  \frac{\left(dx^i\right)^2}{|dx^0|} 
- V\left(x^0,x^i\right) |dx^0|, \label{eq.finsler_quad}
\end{eqnarray}
in standard Cartesian coordinates with $i=1,2, \cdots , n$. 
We assume the foliation defined by $s=x^0$ on $M = \mathbb{R} \times \mathbb{R}^n$. 
The area of the intersection calculated by taking an appropriate Riemannian structure 
is given by 
\begin{eqnarray}
\pp \Delta \Sigma_s \cap D_p \pp_R = {\cal V}_n(r), \quad r = \sqrt{\frac{2 \hbar \D s}{m}},
\end{eqnarray}
where ${\cal V}_n(r)$ is an Euclidean 
volume of $n$-dimensional sphere with radius $r$ (FIG.\ref{fig2}),
and we took $\hbar$ as a natural unit when defining the indicatrix and indicatrix body. 
The contribution from the potential term vanishes in the limit of $\D s \rightarrow 0$.  
Considering two domains 
$\D \Sigma_{s} \cap D_p$ and $d x^1 \cdots d x^n $, 
we find from (\ref{eq.area2}),
\begin{eqnarray}
d \Sigma_{s} = \pp d x^1 \cdots d x^n \pp_F 
=\left(\frac{m}{2 i \hbar \pi \D s}\right)^{n/2} d x^1 \cdots d x^n ,
\end{eqnarray}
with infinitesimally small $\D s$. 
The overall constant factor $\omega$ 
is set appropriately to meet the normalisation condition. 

For the case of harmonic oscillator, 
we can do the calculation in a quite simple way using the properties of Finsler path integral. 
The Finsler function is,
\begin{eqnarray}
 F(t,x,dt,dx)=\frac{m}{2}\frac{(dx)^2}{|dt|}
-\frac{m\omega^2}{2} x^2 |dt|
\end{eqnarray}
in the standard Cartesian coordinates.
Consider a coordinate transformation,
\begin{eqnarray}
 \left\{
\begin{array}{l}
 x=A\cos{(\omega \tau)}\\
 \tau=t-t'
\end{array}
 \right.
\end{eqnarray}
then,
\begin{eqnarray}
& F(\tau,A,d\tau,dA) = \displaystyle\frac{m}{2}
\cos^2{(\omega\tau)}\frac{(dA)^2}{|dt|}
 -dG(\tau,A), \nonumber \\
& G(\tau,A):=\displaystyle\frac{m\omega}{4} 
    A^2\sin{(2\omega \tau)}{\rm sgn}(d\tau).
\end{eqnarray}
Next, consider the reparametrisation of the time variable by,
\begin{eqnarray}
 s=\frac{1}{\omega}\tan{(\omega \tau)}, \quad
  ds=\frac{d\tau}{\cos^2{(\omega \tau)}}.
\end{eqnarray}
We get,
\begin{eqnarray}
 F(s,A,ds,dA)&=&
  \frac{m}{2}
  \frac{(dA)^2}{|ds|}
-dG(\tau(s),A) \label{eq.fins-harmonic}
\end{eqnarray}
For calculational simplicity, we choose the parametrisation $s$, instead of $t$. 
Then the integration is carried out immediately, 
since the second term in (\ref{eq.fins-harmonic}) 
is exact and cancel through the intermediate leaves, 
and the first term is the same as for a free particle. 
As mentioned previously, the measure part is not affected by the potential. 
We have, 
\begin{eqnarray}
&&\mathscr{U}[\Sigma_{s''},\Sigma_{s'}]= \lim_{N \rightarrow \infty}
d \Sigma_{s'} 
\int d\Sigma_{s_1}\int d\Sigma_{s_2} \cdots \int d\Sigma_{s_{N-1}} 
\prod_{j=1}^{N-1}
\left(\frac{m}{2\pi i \hbar \Delta s_j}\right)^{\frac{1}{2}} \nonumber \\
&&\times  
\exp \left[ \frac{i}{\hbar} 
\sum_{k=0}^{N-1}
\int_{ \gamma_{s_{k-1}}^{s_{k}}}
 \frac{m}{2} \frac{(dA)^2}{ds}  
-dG(\tau(s),A)
\right] 
\nonumber \\
&&= \lim_{N \rightarrow \infty} 
\left(\frac{m}{2\pi i\hbar \Delta s_0}\right)^\frac12 dA' 
\int dA_1\int dA_2 \cdots \int dA_{N-1} 
\prod_{j=1}^{N-1}  
\left(\frac{m}{2\pi i \hbar \Delta s_{j}}\right)^{\frac{1}{2}} \nonumber \\
&&\times 
\exp \left[ \frac{i}{\hbar} 
\sum_{k=0}^{N-1}
\left\{  
 \frac{m\left(A_{k+1}-A_{k}\right)^2}{2\Delta s_k}  
-G(\tau(s_{k+1}),A_{k+1})+G(\tau(s_{k}),A_{k})
\right\} \right] 
\end{eqnarray}

Here, $\Delta s_{k} = s_{k+1} - s_{k}$, and 
$s_0= s', \, s_N=s'', \, A_0 =A', \, A_N = A''$. 
The integration gives, 
\begin{eqnarray}
\mathscr{U}[\Sigma_{s''},\Sigma_{s'}]
&=&\left(\frac{m}{2\pi i \hbar (s'' - s')}\right)^{\frac12} dA'
\exp \left[\frac{i m }{2 \hbar}\frac{(A''-A')^2}{s'' - s'} \right] \nonumber \label{eq.harmonic} \\
&\times&   
\exp \left[{ -\frac{i }{\hbar}} 
\left( 
G(\tau(s''),A'')-G(\tau(s'),A').
\right) \right] 
\end{eqnarray}
This propagator depends on parametrisation but not on 
local coordinates of the hypersurfaces. 
We transform the coordinates back to $(t,x)$ and consider the 
transformation rules of the propagator 
between different parameterisation from $s$ to $t$. 
Then one obtains the common description of propagator for a 
harmonic oscillator~\cite{Feynman1}, 
\begin{eqnarray}
\mathscr{U}[\Sigma_{t''},\Sigma_{t'}]=
\sqrt{ \frac{m \omega}{2\pi i \hbar \sin{\omega T}}}
\exp \left[ \frac{im\omega}{2\hbar \sin{\omega T}}
\left\{
\cos \omega T (x''^2+x'^2) -2 x''x'
\right\}
\right]dx'. \label{eq.harmonic2}
\end{eqnarray}
The details of the transformation rules of propagators 
are given in the appendix.

We could show that Finsler path integral is coordinate free by calculating the propagator in 
spherical coordinates for the above simple Finsler function (\ref{eq.finsler_quad}). 
For $(3+1)$-dimension, the Finsler measure in spherical coordinate becomes, 
\begin{eqnarray}
d\Sigma_{s} = \left( \frac{m }{2 i \pi \hbar \D s} \right)^{3/2} r^2 
\sin \theta dr d \theta d \varphi, 
\end{eqnarray}
and integration gives the same result. 

Another example is the case for a 
non-relativistic particle on a Riemannian manifold $(Q,g)$. 
The Finsler manifold we consider for this case is $(M, F)$, 
with $M = \mathbb{R} \times Q$ and the Finsler function defined by 
\begin{eqnarray}
{F}\left(x^0,x^i,dx^0,dx^i\right) 
= \frac{m}{2} \frac{ g\left( dx, dx\right) }{|dx^0|}, 
\end{eqnarray}
where $g$ is the Riemannian metric on the 
manifold $Q$, with $i=1,2, \cdots ,n$. 
We assume the foliation defined by $s=x^0$ on $M = \mathbb{R} \times Q$. 
The additional term that corresponds to Jacobian 
which appears during the quantisation by phase space path integral;
usually referred to as Lee-Yang term, 
could be obtained easily by considering the intersection, 
$\pp \Delta \Sigma_s \cap D_p \pp_R = {\cal V}_n\left
(\sqrt{2  \hbar g\left( dx, dx\right)\D s/m}\right), $ 
and the measure becomes, 
\begin{eqnarray}
d\Sigma_{s} =\left(\frac{m}{2 i \hbar \pi \Delta s} \right)^{n/2} 
\sqrt{\det g}~d x^1 \cdots d x^n.
\end{eqnarray}
Therefore, we could derive this term from Lagrangian formalism only. 

\begin{figure*}[htbp]
 \begin{center}
  \includegraphics[width=90mm]{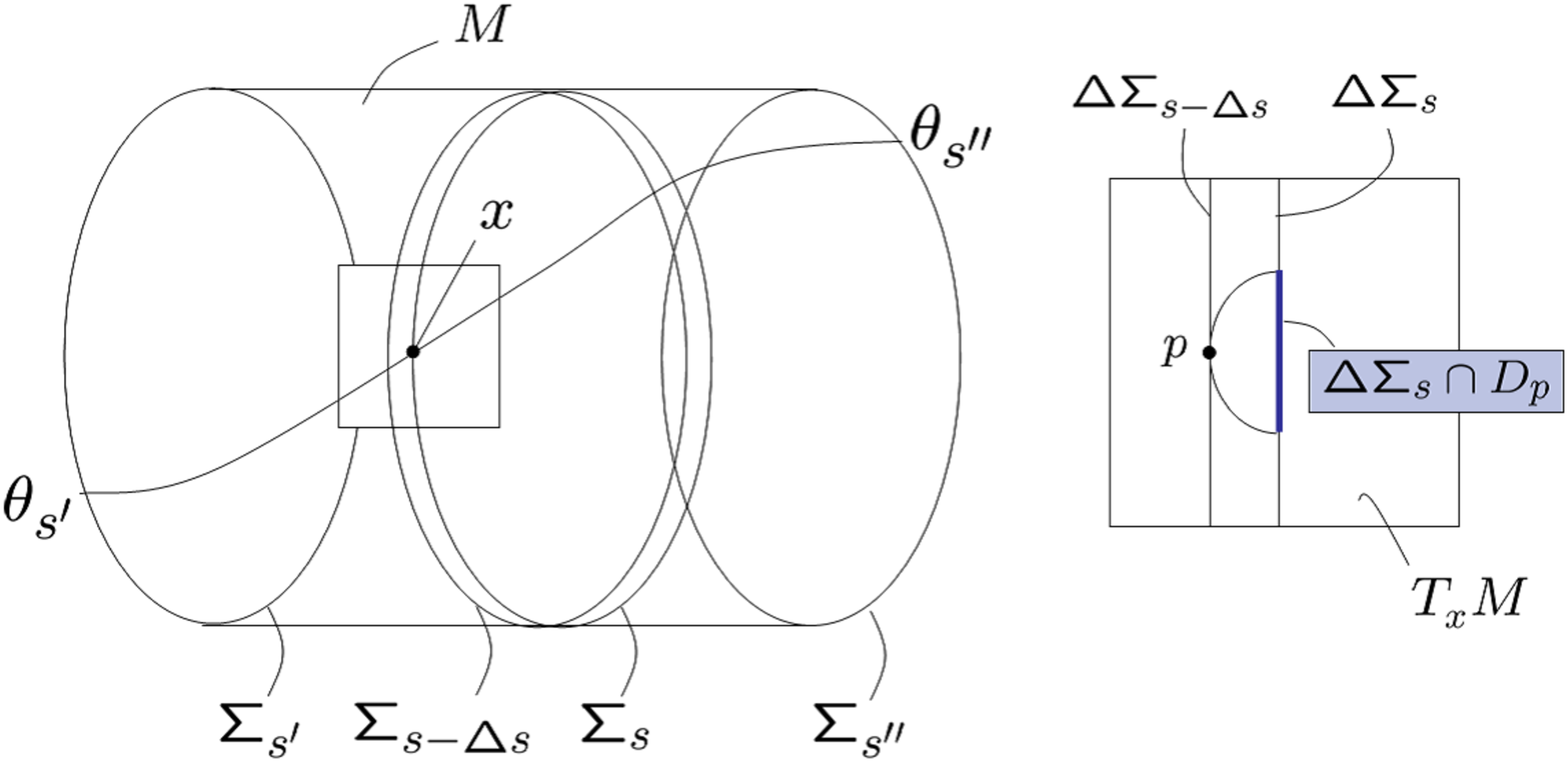}
 \end{center}
  \caption{ }
  \label{fig3}
\end{figure*}
Last example is quantisation of a particle constrained on $S^1$. 
The Finsler manifold is $({\M}, F)$, 
with ${\M} = \mathbb{R} \times S^1_r$, where $S^1_r$ is a circle with radius $r$, 
and the Finsler function defined by 
\begin{eqnarray}
{F}\left(t,\theta,dt,d \theta\right) = \frac{m}{2}\frac{r^2 \left(d \theta\right)^2}{|dt|} .
\end{eqnarray}
We assumed the foliation in a similar way to the previous examples.
The intersection of the indicatrix body 
$D_p$ 
and the hyperplane $\D \Sigma_s$ is a line segment (FIG.\ref{fig3}). 
The measure is, 
\begin{eqnarray}
d\Sigma_s \:= \sqrt{\frac{mr^2}{2\hbar \Delta s}}d \theta .
\end{eqnarray}
Since we need to consider all the geodesics 
in the integrand of (\ref{eq.finslerPI}), 
there are multiple contributions from the paths winding around the cylinder. 
The calculated result coincide with the propagator given in the 
standard text book~\cite{Kashiwa-Ohnuki-Suzuki}. 

These examples, especially harmonic oscillator and particle constrained on $S^1$, shows 
how the calculation complexity could be reduced drastically by using Finsler path integral. 

\section{Discussions} \label{sec.disucussions}
The introduced Finsler path integral is by itself a mathematically sound and 
independent quantisation scheme from canonical quantisation. 
It is coordinate free, and also covariant which means one can choose arbitrary time variable, $s$. 
In principle, the form of Lagrangian needs not be quadratic, 
not even a polynomial. 
The attempt to give the Feynman's path integral a mathematically rigorous 
definition without the use of Hamiltonian formalism was also proposed by 
DeWitt-Morette for the case of a quadratic Lagrangian~\cite{Morette}. 
By their method, the examples we have introduced above could be calculated correctly, 
but it lacks geometrical setting and is not covariant in the sense they have a fixed foliation. 
The Finsler geometrical setting gives essentially a reparametrisation invariant description; 
therefore, it becomes constrained systems. 
The necessary gauge fixing condition corresponds to 
choosing the foliation, equivalently the time variable $s$. 
In the sense that the foliation (or time variable) could be adjusted, 
Finsler path integral is a covariant description. 
The examples we have chosen only permits $s = x^0$ gauge, but for the case such as 
relativistic particle should permit more flexible choice of gauges. 
Since the chosen examples are the well-studied basic ones 
and the results coincide, 
the Finsler path integral may seem a mere reformulation of an old theory. 
However, this reformulation gives us a clear view 
in understanding problems the conventional path integral suffered. 
We have shown that provided with a Lagrangian, one could obtain a correct propagator, 
regardless of the coordinates. 
This is in contrast to canonical quantisation, where there exist various quantum theories, 
depending on the choice of coordinates. 
Not just for known problems, 
but Finsler path integral could be a guide in considering more general physical systems. 
This is a natural prediction since Finsler geometry covers wider 
range of application than the conventional Lagrangian formalism. 
Characteristically, it is capable of expressing irreversible systems and 
hysteresis phenomena, 
and therefore one expects that Finsler path integral could give a sophisticated construction 
to quantisation of these problems. 
Further extension to string theory, 
system of higher-order differential equations and field theory 
also could be considered, and the former two should be 
constructed on a Kawaguchi space, which is a generalisation of 
Finsler geometry. 
The profoundness of the original ideas of path integral and Finsler geometry gives us 
wide varieties of these applications, 
which may shed us some lights on further understanding of 
quantum theory, and possibly lead us to a new discovery. 

\section{Acknowledgments}
We thank Lajos Tamassy for introducing us his work on explicit expression of area 
in Finsler space.
Masahiro Morikawa and Morikawa lab. were always helpful with creative discussions. 
The work is greatly inspired by late Yasutaka Suzuki.                      
This work was supported by YITP of Kyoto University. 
E.T. thanks the Dean of the Science Faculty of Palacky University, 
and grant of the Czech Science Foundation (No. 201/09/0981) for financial support.

\section{Appendix: Transformation rules between propagators with different parametrisation}
Here we show the transformation rules 
of the propagator between different parametrisation, 
which we used for the example of 
harmonic oscillator. For an arbitrary parametrisation $\tau$, 
the Finsler path integral shows that the 
squared absolute value of function $\psi$
over the hypersurfaces are preserved by, 
\begin{eqnarray}
\int_{\Sigma_{\tau''}} d \Sigma_{\tau''} 
\left|\psi_{\Sigma_{\tau''}}\right|^2 = 
\int_{\Sigma_{\tau'}} d \Sigma_{\tau'} 
\left|\psi_{\Sigma_{\tau'}}\right|^2. \label{eq.norm}
\end{eqnarray}
The Finsler measure for the parameter $s$ and $t$ are given by, 
\begin{eqnarray}
 d\Sigma_{s}=\sqrt{\frac{m}{2\pi i\hbar \Delta s}}dA, \quad
 d\Sigma_{t}=\sqrt{\frac{m}{2\pi i\hbar \Delta t}}dx, 
\end{eqnarray}
therefore, substituting it into (\ref{eq.norm}), we see that 
the normalisation condition of $\psi_{\Sigma_t}$ and $\psi_{\Sigma_s}$ must be, 
\begin{eqnarray}
 \int_{\Sigma_s} dA ~|\psi_{\Sigma_s}|^2=1,
  \quad
 \int_{\Sigma_t} dx ~|\psi_{\Sigma_t}|^2=1.
\end{eqnarray}
Now, the relation between the coordinates $x$ and $A$ was 
$x = A \cos (\omega (t-t'))$, therefore, 
$dx=dA\cos (\omega (t-t'))$ 
on the hypersurface $\Sigma_{t}$ 
($t=\mbox{const.}$). 
For the initial hypersurface $\Sigma' := \Sigma_{t'} = \Sigma_{s'}$, 
$dx' = dA'$, so we can take $\psi_{\Sigma_{s'}}=\psi_{\Sigma_{t'}}$.
However, on the final hypersurface $\Sigma'' := \Sigma_{t''} = \Sigma_{s''}$, 
we need to consider 
\begin{eqnarray}
  1=\int_{\Sigma_{s''}} dA'' ~|\psi_{\Sigma_{s''}}|^2
   =\int_{\Sigma_{t''}} dx''\frac{1}{\cos{(\omega T)}} ~|\psi_{\Sigma_{s''}}|^2, 
\end{eqnarray}
which shows, 
$\sqrt{1/\cos{\omega T}}\psi_{\Sigma_{s''}}=\psi_{\Sigma_{t''}}$ 
up to order $1$ phase factor. 
Using the evolution equation 
\begin{eqnarray}
 \psi_{\Sigma_{s''}}= \int_{\Sigma_{s'}}\mathscr{U}[\Sigma_{s''},\Sigma_{s'}]
\psi_{\Sigma_{s'}},\quad
\psi_{\Sigma_{t''}}= \int_{\Sigma_{s'}}\mathscr{U}[\Sigma_{t''},\Sigma_{t'}]
\psi_{\Sigma_{t'}},
\end{eqnarray}
the transformation condition between the propagators of different parametrisation becomes, 
\begin{eqnarray}
\mathscr{U}[\Sigma_{t''},\Sigma_{t'}]= 
\sqrt{\frac{1}{\cos{\omega T}}} \mathscr{U}[\Sigma_{s''},\Sigma_{s'}]
\label{eq.Ut-to-Us}.
\end{eqnarray} 



\begin{thebibliography}{7}
\providecommand{\natexlab}[1]{#1}
\providecommand{\url}[1]{\texttt{#1}}
\providecommand{\urlprefix}{URL }
\expandafter\ifx\csname urlstyle\endcsname\relax
  \providecommand{\doi}[1]{doi:\discretionary{}{}{}#1}\else
  \providecommand{\doi}[1]{doi:\discretionary{}{}{}\begingroup
  \urlstyle{rm}\url{#1}\endgroup}\fi
\providecommand{\bibinfo}[2]{#2}

\bibitem[{Feynman(1948)}]{Feynman1}
\bibinfo{author}{R.~P. Feynman}, \bibinfo{title}{Space-Time Approach to
  Non-Relativistic Quantum Mechanics}, \bibinfo{journal}{Rev. Mod. Phys.}
  \bibinfo{volume}{20} (\bibinfo{year}{1948}) \bibinfo{pages}{367--387}.

\bibitem[{Tamassy(1993)}]{Tamassy1}
\bibinfo{author}{L.~Tamassy}, \bibinfo{title}{AREA AND CURVATURE IN FINSLER
  SPACES}, \bibinfo{journal}{Rep. Math. Phys.} \bibinfo{volume}{33}
  (\bibinfo{year}{1993}) \bibinfo{pages}{233--239}.

\bibitem[{Cartan(1934)}]{Cartan1}
\bibinfo{author}{E.~Cartan}, \bibinfo{title}{LES ESPACES DE FINSLER},
  \bibinfo{journal}{Conf\'erence faite le 18 mai 1934}  (\bibinfo{year}{1934})
  \bibinfo{pages}{1385}.

\bibitem[{Chern and Shen(2005)}]{Chern2}
\bibinfo{author}{S.~S. Chern}, \bibinfo{author}{Z.~Shen},
  \bibinfo{title}{RIEMANN-FINSLER GEOMETRY}, \bibinfo{publisher}{World
  Scientific}, \bibinfo{year}{2005}.

\bibitem[{Busemann(1947)}]{Busemann2}
\bibinfo{author}{H.~Busemann}, \bibinfo{title}{Intrinsic Area},
  \bibinfo{journal}{Ann. of Math.} \bibinfo{volume}{48} (\bibinfo{year}{1947})
  \bibinfo{pages}{234--267}.

\bibitem[{Kashiwa et~al.(1997)Kashiwa, Ohnuki, and
  Suzuki}]{Kashiwa-Ohnuki-Suzuki}
\bibinfo{author}{T.~Kashiwa}, \bibinfo{author}{Y.~Ohnuki},
  \bibinfo{author}{M.~Suzuki}, \bibinfo{title}{Path Integral Methods},
  \bibinfo{publisher}{Oxford University Press}, \bibinfo{year}{1997}.

\bibitem[{Morette(1951)}]{Morette}
\bibinfo{author}{C.~Morette}, \bibinfo{title}{On the definition and
  approximation of Feynman's path integral}, \bibinfo{journal}{Phys. Rev.}
  \bibinfo{volume}{81} (\bibinfo{year}{1951}) \bibinfo{pages}{848--852}.

\end{thebibliography}
\end{document}